\documentstyle[preprint,aps]{revtex}
\begin{document}
\draft
\preprint{CLNS 96/1432}
\title{Qualitative Aspects of Polarization Distributions In Excited 
Heavy Hadron Productions}
\author{Chi--Keung Chow}
\address{Newman Laboratory of Nuclear Studies, Cornell University, Ithaca,
NY 14853.}
\date{\today}
\maketitle
\begin{abstract}
Within the context of flux tube models, heavy quark fragmentation takes 
place through the breaking of flux tubes with the production of a 
(di)quark-anti(di)quark pair.  
It is found that the (di)quark produced are more likely to be found in 
an $L_z=0$ state.  
This naturally leads to an supression of the polarization distribution 
parameters $w_{3/2}$ and $\tilde w_1$ for $(D_1, D_2^*)$ and $(\Lambda_{c1}, 
\Lambda_{c1}^*)$ production respectively.  
The corresponding parameter $w_1$ for $(\Sigma_c, \Sigma_c^*)$ production, 
however, is not suppressed, in agreement with the CLEO results but not 
the DELPHI one.  
Implications on the measurements of $\Lambda_Q$ polarizations are discussed.  
\end{abstract}
\pacs{}
\narrowtext
The production of excited heavy hadrons has been studied in Ref.~\cite{1}.  
It was observed that, due to heavy quark symmetry and parity conservation 
of the strong interaction, the relative production probabilities of the 
different helicity states of the light degrees of freedom can only depend 
on the absolute magnitude of $j$, the helicity of the light degrees of 
freedom along the production axis, but not on its sign. 
As a result, the relative production probabilities of the $(D_1, D_2^*)$ 
system is controlled by a single parameter $w_{3/2}$, which is defined to 
be the probability that $j$ has its maximal value $3\over2$.  
A similarly defined parameter $w_1$ controls the production of the $(\Sigma_c, 
\Sigma_c^*)$ system, while another parameter $A$ describes the likelihood of 
the production of a spin-1 diquark instead of a spin-0 one, which translates 
into the probability of producing a $\Sigma_c$ or $\Sigma_c^*$ instead of a 
$\Lambda_c$.  
The framework has been subsequently extended to describe the excited 
$\Lambda$ resonances $(\Lambda_{c1}, \Lambda_{c1}^*)$ \cite{2}, with two 
more parameters $(B, \tilde w_1)$ defined in analogy to $(A, w_1)$ for the 
$(\Sigma_c, \Sigma_c^*)$ system.  

It must be emphasized that Ref.~\cite{1} and \cite{2} are parametrizations 
rather than predictions of the fragmentation processes in the sense that 
they did not attempt to predict (or explain) the experimental values of 
the $w$'s.  
By measuring the angular distribution of the decay products, ARGUS \cite{3,4} 
and CLEO \cite{5,6} have measured $w_{3/2}$.  
They found $w_{3/2}$ to be small (best fit $w_{3/2}$ is $-0.30$, which is in 
the unphysical region; restricting to the physical region $0 \le w_{3/2}\le 1$ 
gives $w_{3/2}\sim 0$), meaning that transverse polarization is preferred to 
longitudinal.  
There is no theoretical understanding of why $w_{3/2}$ is so small.  
In the case of excited $B_c$ production, one can calculate $w_{3/2}$ by 
perturbative QCD \cite{7,8}, and the result $w_{3/2}=29/114\sim0.24$ is 
indeed on the small side. 
But it is dangerous to simply carry the result over to the $(D_1, D_2^*)$ 
where non-perturbative QCD effects are dominant.  
In the baryon sector, there is no measurement for $\tilde w_1$ yet, while 
the $w_1$ measurements by CLEO and DELPHI yielded inconsistent results.  
DELPHI\footnote{The DELPHI analysis use $\Sigma_b^{(*)}$ production from 
$Z^0$ decays, not $\Sigma_c^{(*)}$.  
By heavy quark symmetry, however, $w_1$ should be the same for both cases.} 
obtained a small value for $w_1$ (best fit $w_1$ is $-0.36$, which is also 
unphysical; restricting to the physical region $0 \le w_1\le 1$ again 
gives $w_1\sim 0$) \cite{a,d} while the CLEO result is consistent with an 
isotropic polarization ($w_1=0.71\pm0.13$ while an isotropic distribution 
gives $w_1={2\over3}$) \cite{b}.  

Intuitively the helicity of the light degrees of freedom can be viewed as 
the sum of two different contributions.  
One is the spin of the ``brown muck'' $\vec S$, which is 0 for $\Lambda$ 
type baryons, $1\over2$ for mesons and 1 for $\Sigma$ type baryons.  
Then this ``brown muck'' may orbit around the heavy quark with orbital 
angular momentum $\vec L$, giving an additional contribution to the 
helicity.  
We will see that, if heavy quark fragmentation can be understood as 
breaking of color flux tubes as suggested by the Lund models \cite{9}, the 
Artru--Mennessier model \cite{10} and the UCLA model \cite{c}, then the 
orbital angular momentum naturally prefers a transverse polarization, 
explaining the smallness of $w_{3/2}$ and predicts a small $\tilde w_1$; 
$w_1$, on the other hand, is not required to be small, {\it i.e.}, the CLEO 
numbers are prefered.  

Due to the non-abelian nature of QCD, the color field is expected to be 
confined into tube-like regions (flux tubes) by the tri-gluon coupling.  
The flux tubes have constant tension and ends at colored objects like quarks 
or diquarks.  
Such a picture is supported by Regge phenomenology, quarkonium spectroscopy, 
bag models and lattice QCD calculations.  
For concreteness, let's study the process $Z^0\to c\bar c$ and the subsequent 
fragmentation and hadronization.  
Just after the $Z^0$ decay, both quarks are in general very off-shell and 
will fragment by the emission of hard gluons, which is governed by 
perturbative QCD.  
Eventually such gluon bremsstrahlung will bring the off-shell energy down 
to $\Lambda_{\rm QCD}$ scale, and non-perturbative QCD will be important.  
This is when the flux tube model become a reasonable description of the 
dynamics.  
In a coordinate system in which the $c$ quark travel along the positive 
$z$-axis in the $c\bar c$ center of mass frame ($Z^0$ rest frame if the 
momenta carried away by the hard gluons are negligible), there will be a 
flux tube lying along the $z$-axis joining the two quarks.  
The flux tube will be characterized by a constant linear energy density 
(tension) $\kappa\sim 0.2 ({\rm GeV})^2$, which leads to a linear potential 
between the quarks.  
For a long flux tube, it will be energetically favorable to break the flux 
tube by the production of a quark-antiquark pair (or a diquark-antidiquark 
pair) to shorten the flux tube and hence reduce the energy stored in the 
flux tube.  
This is the QCD analog of pair creation in a strong electric field and is 
completely non-perturbative in nature.  

If we ignored the finite thickness of the flux tube, the pair-created 
quark-antiquark pair will be produced right on the $z$-axis, where the flux 
tube is.  
Moreover, due to the tension of the flux tube, the antiquark will be linked 
by the flux tube to the $c$ quark and move towards the $c$ quark, {\it i.e.,} 
along the positive $z$ direction.  
In general there will also be transverse momenta, which has a gaussian 
distribution centered at zero, but let us ignore that for a moment.  
Then all colored objects (quarks and flux tubes) are on the $z$-axis, and 
the system has a rotational symmetry about the $z$-axis.  
As a result, the $z$-component of the orbital angular momentum $L_z$ is 
conserved.  
Since the system starts out with vanishing $L_z$ (nothing is orbiting), the 
final hadron must have $L_z=0$.  
Notice that $L^2$ of the final hadron is not necessarily equal to zero, as 
the system does not have a spherical symmetry and hence $L^2$ is not 
conserved.  

An alternative way of seeing the same result is to consider the wave 
function of the antiquark in the relative momentum space\footnote{An 
similar argument in the relative position space also holds analogously.}.  
Since the light antiquark is moving towards to $c$ quark, it means that the 
relative momentum $\vec p$ of the light antiquark with respect to the $c$ 
quark is in the positive $z$ direction.  
In other words, the wave function of the antiquark in the momentum space is 
a wave packet peaked at some point $\vec p_0$ on the positive $z$ axis.  
We do not know the exact shape of the wave function, but the rotational 
symmetry about the $z$ axis mandates that the wave function can depend only  
on $p_z$ and $\sqrt{p_x^2 + p_y^2}$, but not the azimuthal angle $\theta$.  
As a result, the expectation value of $L_z=i \partial_\theta$ vanishes.  
Notice that our argument holds even if the spread of the wave function is 
large with respect to $|\vec p_0|$ and the wave function is non-vanishing 
even for points off the $z$-axis.  
As long as the wave function is azimuthally symmetric, $L_z$ has to 
vanish.  
Hence we see that $L_z$ in heavy meson production vanishes if transverse 
momenta are negligible.  
The polarization of the intrinsic spin of the antiquark $\vec S$, on the 
other hand, is not constrained in any way as long as it is cancelled by 
that of the quark produced at the same time.  
Since the flux tube models do not have a preference over the orientation 
of $\vec S$, we naturally {\it assume\/} it to be isotropic.  
The total helicity of the light degrees of freedom, then, is the sum of 
of $\vec L$, under the constraint $L_z = 0$, and $\vec S$, with an 
isotropic distribution.  

For a heavy meson with orbital angular momentum $L^2$, the possible 
helicity states for the light degrees of freedom range over $j=-L-{1\over2}, 
-L+{1\over2}, \dots, L-{1\over2}, L+{1\over2}$.  
By the conservation of parity, the probability of finding the light degrees 
of freedom with helicity $j$ is the same as that with helicity $-j$.  
We will define $W_j$ the probability of finding the light degrees of freedom 
with helicity either $j$ or $-j$.
The sum of all these probabilities equals to unity, {\it i.e.}, $\sum 
W_j = 1$.  
In particular, for $L=1$, $W_{3/2} = w_{3/2}$ defined in Ref. \cite{1}, 
and $W_{1/2} = 1 - w_{3/2}$.  
The analysis above suggests that $W_{1/2}=1$ and all other $W$'s vanish, 
for all values of $L$.  
In other words, the light degrees of freedom will be in the lowest helicity 
state, and be as transversely polarized as allowed by quantum mechanics.  
In particular, for the $(D_1, D_2^*)$ system, $w_{3/2}=W_{3/2}=0$ seen 
by ARGUS and CLEO \cite{3,4,5,6}.  

Our formalism can be extend to describe the production of heavy baryons as 
well.  
Different flux tube models have different descriptions of the production 
of the diquark-antidiquark pairs.  
In the simplest models, the diquark appears as a single entity, and the 
analysis above can be adopted in a straightforward manner. 
In some other models, like the ``popcorn model'' \cite{11}, the two quarks 
in the diquark are produced in stages.  
The analysis above will be invalidated if the first quark-antiquark pair 
moves off the $z$-axis before the second pair is created, as such 
off-axis configuration will break the azimuthal symmetry.   
It turns out that, however, the first quark-antiquark pair will instead 
slide along the flux tube as ``curtain quarks'' but not wander off the 
flux tube.  
Hence, even in these models, all colored objects still lie on the 
$z$-axis, azimuthal symmetry is preserved, and hence the resultant heavy 
baryon will also have $L_z=0$.  

Since $S=0$ for a $\Lambda$ type diquark, the light degrees of freedom of a 
$\Lambda$ type heavy baryon with orbital angular momentum $L^2$ can have 
helicity $j=-L, \dots, L$.  
Define as before $W_j$ as the probability of finding the light degrees of 
freedom with helicity either $j$ or $-j$.
Then the analysis above suggests that $W_0=1$ and all other $W$'s vanish.  
In particular, for the $(\Lambda_{c1}, \Lambda_{c1}^*)$ system, $W_1= 
\tilde w_1$ as defined in Ref. \cite{2}, and $W_0=1- \tilde w_1$.  
Our analysis then predicts $\tilde w_1$ to be small.  
On the other hand, $S=1$ for a $\Sigma$ type diquark, and the light degrees 
of freedom of a $\Sigma$ type heavy baryon with orbital angular momentum 
$L^2$ can have helicity $j=-L-1, -L, \dots, L, L+1$.  
Since $\vec S$ is supposed to have an isotropic distribution, $S_z$ is 
equally likely to be found in the $+1$, 0, or $-1$ states.  
This gives $W_0={1\over3}$, $W_1={2\over3}$ and all the other $W$'s vanish.  
For the $(\Sigma_c, \Sigma_c^*)$ system, which is {\it not\/} orbitally 
excited, our analysis suggests an isotropic distribution, {\it i.e.}, 
$w_1={2\over3}$ for $w_1$ defined as in Ref. \cite{1}.  
This is in agreement which the CLEO result \cite{b} but not the DELPHI one 
\cite{a,d}.  

Due to heavy quark symmetry, our analysis is obviously also applicable to 
$b$ quark fragmentation as well.  
It can also be easily generalized to other excited heavy hadrons.  
By the $L_z=0$ rule, the light degrees of freedom of excited heavy mesons 
will have $|j|={1\over2}$, excited $\Lambda$ type baryons $|j|=0$, and 
for excited $\Sigma$ type baryons, $|j|$ can either be 0 or 1, with 
probabilities ${1\over3}$ and ${2\over3}$ respectively.  
One exception is the $P$-wave $\Sigma$ type baryon, with the spin of light 
degrees of freedom $\vec s_\ell = \vec L + \vec S = \vec 0$.  
Then there will be only one helicity state and $W_0 = 1$ trivially.  
In general, however, care must be taken to apply our analysis to very 
excited heavy hadrons, as the flux tube models may cease to be good 
descriptions with high excitation energies.  

It is interesting to see the implication of our analysis on the measurements 
of heavy quark polarization in $Z^0\to c\bar c, b\bar b$ processes.  
As discussed in Ref.~\cite{1}, in general (except one special case) all 
polarization information of the heavy quark is lost in the meson sector.  
On the other hand, since the ``brown muck'' of $\Lambda_Q$ is spinless, 
the polarization of the heavy quark should be retained in the baryon sector.  
This effect, however, is modified by the presence of secondary $\Lambda_Q$'s, 
{\it i.e.}, those produced in decays of excited heavy baryons.  
Since the polarizations of these secondary $\Lambda_Q$'s do not necessarily 
align with the initial heavy quark, the overall polarization is diluted.  
It has been shown \cite{2} that, if one only includes secondary $\Lambda_Q$'s  
from the $(\Sigma_Q, \Sigma^*_Q)$ and $(\Lambda_{c1}, \Lambda_{c1}^*)$ 
doublets, the polarization is diluted by the factor 
\begin{equation}
{\cal P} = {1 + {A\over 9}(1+4w_1) + {B\over 9}(1+4\tilde w_1) \over 1+A+B}.  
\end{equation}
The first, second and third terms in both the numerator and the denominator 
correspond to $\Lambda_Q$ produced directly, from $(\Sigma_Q, \Sigma^*_Q)$ 
decays and and from $(\Lambda_{c1}, \Lambda_{c1}^*)$ decays respectively.  
With $w_1={2\over3}$, $\tilde w_1=0$ and both $A$ and $B$ assuming the 
default Lund value 0.45 \cite{9}, it is found that ${\cal P} = 0.65$.  
For comparison, ${\cal P}=0.58$, $0.72$ and $0.79$ for $w_1 = \tilde w_1 
=0$, $2\over3$ and 1 respectively.  
In the standard model, one find that the $b$ and $c$ quarks produced by 
$Z^0$ decay is partially polarized with ${\cal P}_b = -0.94$ and 
${\cal P}_c = -0.67$.\footnote{The definition of polarization used here 
differs from that in Ref.~\cite{1} by a negative sign to facilitate 
comparison with the ALEPH results.}  
Hence our analysis predicts ${\cal P}_{\Lambda_b} = {\cal P}_b {\cal P} = 
-0.61$ and ${\cal P}_{\Lambda_c} = {\cal P}_c {\cal P} = -0.44$.  
Note that this is two standard deviations away from both the ALEPH result 
\cite{12} ${\cal P}_{\Lambda_b} = -0.23{+0.24 \atop -0.20}$(stat.)$+0.08
\atop -0.07$(syst.) and the recent DELPHI preliminary result \cite{13} 
${\cal P}_{\Lambda_b} = -0.08{+0.35 \atop -0.29}$(stat.)$+0.18
\atop -0.16$(syst.).  
This should not be interpreted as the failure of our analysis, since the 
ALEPH and DELPHI central values $-0.23$ and $-0.08$ correspond to 
${\cal P} = 0.24$ and $0.08$ respectively, which are not achievable for any 
choice of $w_1$ and $\tilde w_1$ anyway.  
Instead, this probably means that effects of other resonances, like the 
$P$-wave $\Sigma_Q$'s and $D$-wave $\Lambda_Q$'s, are not negligible.  
Also the hypothesis that $A=B=0.45$ has not yet been tested.  
In fact, DELPHI preliminary results ($1<A<2$ with large uncertainty) 
\cite{d} suggests that more $\Sigma^{(*)}_Q$ are produced than the Lund 
model expects, and hence the depolarization is more severe.  
Evidently more accurate measurements on the parameters $A$, $B$ and the 
$w$'s are necessary to clarify the situation.  

The leading correction to our analysis comes from the transverse momenta 
acquired by the (di)quark-anti(di)quark pair.  
These transverse momenta have a random (gaussian) distribution with 
$\langle p_\perp \rangle = 0$ and $\langle p_\perp^2 \rangle \sim (0.3
{\rm GeV})^2$.  
Our analysis is a good description only if $\langle p_z^2 \rangle\gg 
\langle p_\perp^2 \rangle$, {\it i.e.}, the produced 
(di)quark-anti(di)quark pair move essentially in the $z$ direction to 
preserve the azimuthal symmetry.  
Since both of these are governed by non-perturbative QCD, they should both 
be of order $\Lambda_{\rm QCD}$, and it is not obvious that 
$\langle p_z^2 \rangle$ should be much larger than 
$\langle p_\perp^2 \rangle$.  
Physically, however, we do expect $\langle p_z^2 \rangle = {1\over2} 
\langle p_\perp^2 \rangle$ in the absence of the flux tube, and hence, 
under the tension of the flux tube, $\langle p_z^2\rangle$ to be at least 
as large as ${1\over2}\langle p_\perp^2 \rangle$.  
Consequently, we expect the probability of having orbitally excited hadrons 
with non-vanishing $L_z$ to be suppressed by $O({1\over2}
\langle p_\perp^2 \rangle / \langle p_z^2\rangle)$.  

We conclude that, in flux tube models, orbitally excited heavy hadrons 
tend to have $L_z=0$.  
This gives a natural explanation of the small observed value of $w_{3/2}$, 
and a small value is predicted for $\tilde w_1$ but not $w_1$.  
A large value of $\tilde w_1$ (for an isotropic distribution $\tilde w_1 =
{2\over3}$) would be fatal to our scheme, while a small $w_1$ will mean 
that there are physics not captured by the flux tube model to make the 
diquark spin $\vec S$ anisotropic.  
It is expected that the next round of experiments will resolve the 
controversy on $w_1$ and possibly measure $\tilde w_1$ as well, putting 
our prediction to test.  
Our analysis is qualitative in nature; our ignorance of the $\langle p_z^2 
\rangle$ prevents us from making quantitative predictions.  
On the other hand, it is important to note that our analysis is entirely 
non-perturbative in nature.  
This complements the perturbative calculations in Ref. \cite{7,8} and 
suggests that the suppression of high helicity states is a genuine 
consequence of QCD, not that of the perturbative approximation.  

\acknowledgments
This paper is inspired by Adam Falk's talk at Cornell \cite{14}.  
I am grateful to David Cassel, John Elwood, Adam Falk and John Yelton for 
valuable discussions.  
The work is supported in part by the National Science Foundation.

\end{document}